# Achieving Approximate Soft Clustering in Data Streams

Vaneet Aggarwal
AT&T Labs - Research
email: vaneet@research.att.com

Shankar Krishnan
AT&T Labs - Research
email: krishnas@research.att.com

**Abstract**

In recent years, data streaming has gained prominence due to advances in technologies that enable many applications to generate continuous flows of data. This increases the need to develop algorithms that are able to efficiently process data streams. Additionally, real-time requirements and evolving nature of data streams make stream mining problems, including clustering, challenging research problems.

In this paper, we propose a one-pass streaming soft clustering (membership of a point in a cluster is described by a distribution) algorithm which approximates the "soft" version of the k-means objective function. Soft clustering has applications in various aspects of databases and machine learning including density estimation and learning mixture models. We first achieve a simple pseudo-approximation in terms of the "hard" k-means algorithm, where the algorithm is allowed to output more than k centers. We convert this batch algorithm to a streaming one (using an extension of the k-means++ algorithm recently proposed) in the "cash register" model. We also extend this algorithm when the clustering is done over a moving window in the data stream.

## 1 Introduction

The problem of clustering a group of data items into similar groups is one of the most widely studied research problems with applications in databases, machine learning and computational geometry. Given a set of points and pairwise distance (or similarity) between the points, clustering algorithms divide the points into sets such that points in each set are "close" or "similar" with respect to an objective function. Clustering problems arise in two main flavors - *hard clustering*, where each point's membership is exclusively to a single cluster, and *soft clustering*, where membership of a point in a cluster is described by a distribution.

Often, clustering problems arise in a geometric setting, where the data items are points in high-dimensional Euclidean space. In such a setting, it is natural to define the distance between two points as the Euclidean distance between them. In this paper, we will assume this setting for the clustering problem. One of the most popular definitions for clustering is the *k-means* problem which is defined as follows. Given an integer $k$ and a set of $n$ data points $\mathcal{X} \in \mathbb{R}^d$, the objective of k-means problem is to give $k$ centers $\mathcal{C}$ so as to minimize the objective function

$$\Phi = \sum_{x \in \mathcal{X}} \min_{c \in \mathcal{C}} d(x, c)^2, \quad (1.1)$$

where $d(x, c) = \|x - c\|$.

Estimating parameters of a distribution from sampled data is one of the oldest and most general problems of statistical inference. Given a number of samples, one needs to choose a distribution that best fits the observed data. While traditionally theoretical analysis in the statistical literature has concentrated on rates (e.g., minimax rates), in recent years other computational aspects of this problem, especially as dependence on dimension of the space, have attracted attention. This effort has been particularly directed at the family of Gaussian Mixture models due to their simple formulation and widespread use in several applications spanning databases, computer vision, and machine learning. There are strong connections between learnability of Gaussian mixtures and clustering [7, 9]. In this context, clustering appears in its "soft" form.

In soft clustering, each data point is assigned to several clusters partially. For each point $x$ we have a coefficient giving the degree of being in the $i^{th}$ cluster $u_i(x)$. Usually, the sum of those coefficients for any given $x$ is defined to be 1, $\sum_{i=1}^{k} u_i(x) = 1$. The objective of soft k-means problem is to give $k$ centers $\mathcal{C}$ so as to minimize the potential function,

$$\Phi = \sum_{x \in \mathcal{X}} \sum_{i=1}^{k} u_i(x) d(x, c_i)^2. \quad (1.2)$$

With the explosive growth of financial, social and scientific data sources, it becomes increasingly important to design clustering algorithms which can process the data in the streaming fashion. In the data stream model of computation, the points are read in a sequence and we desire to compute a function, clustering in our case, on the set of points seen so far. This is called the *cash register* model. In typical applications, the total volume of data is very large and can not be stored in its entirety. Another model for streaming is called the *moving window* model, where the function is computed only on L most recent points seen in the stream. This is, typically, of more interest in a practical setting. However, with both insertion and deletion of points, algorithm design gets more challenging.

## 1.1 Our Contributions

Our main result establishes the relationship between hard and soft clustering. For a particular form of soft clustering, called in the literature as *fuzzy k-means* (or fuzzy c-means), we show that k-means approximates it by a factor $O(f(k))$ ($f(k)$ is a polynomial in k and its precise form depends on a parameter in the definition of fuzzy k-means). This result, coupled with the $\Theta(\log k)$ approximation result for k-means of Arthur and Vassilvitskii [4], we obtain the first approximation algorithm for fuzzy k-means.

Ailon *et al.* [3] extend the k-means++ algorithm to a streaming algorithm (in the cash register model). A secondary result in our paper is to adapt their algorithm into a streaming version in the *moving window* model. A natural consequence of our work is a streaming algorithm for soft clustering in both streaming models.

## 2 Previous Work

One of the most popular heuristic algorithms for k-means is Lloyds algorithm [20], which initially chooses k centers randomly. For each input point, the nearest center is identified. Points that choose the same center belong to a cluster. New centers are calculated for the clusters by computing the centroid of points within a cluster. This process is repeated until no changes occur. It is easy to show that the cost function does not increase during any iteration. Hence, this algorithm converges to a local minimum. Its main attractiveness is its simplicity and speed. However, there is no guarantee on the quality of the obtained solution [18].

The fastest exact algorithm for the k-means clustering problem was proposed by Inaba *et al.* [17]. They observed that the number of Voronoi partitions of k points in $\mathbb{R}^d$ is $O(n^{kd})$ and so the optimal k-means clustering could be determined exactly in time $O(n^{kd+1})$. They also proposed a randomized $(1+\epsilon)$-approximation algorithm for the 2-means clustering problem with running time $O(n/\epsilon^d)$.

The k-means problem is known to be NP-hard even for $k=2$ [13]. Matousek [21] gave the first PTAS for this problem, with running time polynomial in n for a fixed k and d. Kanungo *et al.* [18] proposed an $O(n^3\epsilon^{-d})$ algorithm that is $(9+\epsilon)$-competitive by adapting the k-median algorithm of Arya *et al.* [5]. Har-Peled and Mazumdar [16] propose a $(1+\epsilon)$-approximate solution to the k-means problem with running time $O(n+k^{k+2}\epsilon^{-(2d+1)k}\log^{k+1}n\log^k(1/\epsilon))$. For fixed k and d, they achieve linear running time. The algorithm uses a coreset construction by sampling in an exponential grid. Kumar *et al.* [19] propose a simple $(1+\epsilon)$-approximation scheme with a running time of $O(2^{(k/\epsilon)^{O(1)}}dn)$, for a fixed k. Their idea is to recursively approximate the centroid of the largest remaining cluster by trying all subsets of constant size from a sample followed by pruning sufficient points from this large cluster.

Mettu and Plaxton [22] propose a technique called successive sampling to achieve a constant factor approximation for the k-median problem. This idea was adapted independently by Ostrovsky *et al.* [23] and Arthur and Vassilvitskii [4] for the k-means problem. The main idea of both these results is to choose the initial centers for Lloyd's algorithm carefully using a clever sampling technique. Ostrovsky *et al.* [23] achieve a constant factor approximation provided the input satisfies an $\epsilon$-*separated* condition. Arthur and Vassilvitskii [4], however, do not make this assumption and achieve a $O(\log k)$-competitive algorithm.

Soft clustering has applications in various applications spanning databases, statistical inference and machine learning. The two most popular versions of soft clustering are fuzzy k-means [14, 6] and Expectation Maximization (EM) [12] algorithms. At a high level, both algorithms are rather similar, performing a two-step iterative optimization until convergence. The first step, called the expectation (E) step, is an assignment of each data point to clusters or density models as a distribution, and the second step, called the maximization (M) step, re-estimates the clusters based on the current assignments. Just like Lloyd's algorithm, the iterative optimization procedure may result in a local optimum. Relatively little is known about these methods from a theoretical point of view. The problem of giving an approximation algorithm to the fuzzy k-means problem is considered open [11].

Guha *et al.* [15] provide a streaming algorithm for the k-median problem. In particular, they propose a simple divide and conquer strategy to give a constant-factor, single-pass approximation in time $\tilde{O}(nk)$ and sublinear $O(n^\alpha)$ space for constant $\alpha > 0$. Charikar *et al.* [8] gave a constant-factor, single-pass k-Center algorithm using $O(nk\log k)$ time and $O(k)$ space. Recently, Ailon *et al.* [3] combined the results of Guha *et al.* [15] and Arthur and Vassilvitskii [4] to propose a $O(\log k)$ factor approximation to the k-means problem. From a practical point of view, Ackermann *et al.* [1] provide a non-uniform sampling approach to obtain small coresets from the data stream to solve the k-means problem.

## 3 Problem Definition and Known Results

### 3.1 k-means problem

These centers (or cluster centers) define a clustering - all the points closest to a center than to any other center define a cluster. Finding an exact solution to the k-means problem is NP-hard. A well known algorithm called "Lloyd's algorithm" [20] is an algorithm that is guaranteed to find a local optimal solution to the problem, which can often be quite poor.

### 3.2 k-means++ algorithm

The authors of [4] proposed a way of initializing k-means by choosing random starting centers with certain probabilities which give a $\Theta(\log k)$-competitive algorithm to k-means problem with a running time of $O(nkd)$. The initial seeding of k-means++ is described in the following algorithm.

1. Choose an initial center $c_1$ uniformly at random from $\mathcal{X}$.

2. Choose the next center $c_i$ selecting $c_i = x' \in \mathcal{X}$ with probability $\frac{D(x')^2}{\sum_{x \in \mathcal{X}} D(x)^2}$, where $D(x)$ is the shortest distance from a data point $x$ to the closest center we have already chosen.

3. Repeat Step 2 until we have chosen a total of $k$ centers.

### 3.3 k-means# Algorithm

The authors of [3] extended the kmeans++ algorithm to give an algorithm that provides $O(k \log k)$ centers to yield $O(1)$ competitive strategy for k-means with constant probability. This algorithm chooses $3 \log k$ centers randomly in the first round. Further, $3 \log k$ centers are chosen in step 2 of k-means++ and is repeated $(k-1)$ times as in k-means++ algorithm. Since the guarantees are with constant probability, the algorithm needs to be repeated large enough times to get better guarantees. For instance, the authors of [3] repeat the algorithm $O(\log n)$ times to get a non-competitive solution with probability at-most $O(1/n)$.

### 3.4 Streaming k-means

A streaming version of k-medians was provided in [15]. This idea was used by the authors of [3] to provide a streaming version of k-means. A multi-level algorithm is used for a given memory order $n^\alpha$. In all but the last level, $n^\alpha$ data points are compressed to $O(k \log k)$ using k-means# algorithm (using best run of $O(\log n)$ trials of k-means#). In the last level, $n^\alpha$ data points will be compressed to $k$ points using k-means++ algorithm. The guarantees using this algorithm can be summarized in the following theorem.

**Theorem 3.1** ([3]). *If there is access to memory of size $M = n^\alpha$ for some fixed $\alpha > 0$, then for sufficiently large $n$ the best application of the multi-level scheme described above is obtained by running $r = O(\log(n/M)/\log(M/k \log k))$ levels (which is constant), and choosing the repeated k-means# for all but the last level, in which k-means++ is chosen. The resulting algorithm is a randomized streaming approximation to k-means, which is $O(\log k)$-competitive. Its running time is $O(dnk^2 \log n \log k)$.*

### 3.5 Soft k-means

In hard clustering, each data point is assigned to its closest center. However, in soft clustering, each data point is assigned to several clusters partially. For each point $x$ we have a coefficient giving the degree of being in the $i^{th}$ cluster $u_i(x)$. Usually, the sum of those coefficients for any given $x$ is defined to be 1, $\sum_{i=1}^{k} u_i(x) = 1$. The centroid of a cluster is the mean of all points, weighted by their degree of belonging to the cluster, or

$$c_i = \frac{\sum_x u_i(x) x}{\sum_x u_i(x)}. \tag{3.1}$$

We will assume following [6] that,

$$u_i(x) = \frac{1}{\sum_j (\frac{d(c_i,x)}{d(c_j,x)})^{2/m}}. \tag{3.2}$$

The objective of soft k-means problem is to give k centers $\mathcal{C}$ so as to minimize the potential function,

$$\Phi = \sum_{x \in \mathcal{X}} \sum_{i=1}^{k} u_i(x) d(x, c_i)^2. \tag{3.3}$$

For $m = 1$, this is equivalent to normalizing the coefficient linearly to make their sum 1. For $m \to 0$, the cluster centers approach k-means centers. We will assume that $0 < m < 1$. For a given number of clusters k, soft clustering is usually done using EM algorithm which can be defined as follows.

1. Choose k centers at random.

2. Repeat until the algorithm has converged :

    (a) For each point x, compute $u_i(x)$.
    (b) Compute the centroid $c_i$ for each cluster.

## 4 k-means as an approximation for soft k-means

In this section, we will show that choosing the centers obtained by k-means give $O(k^{m/(1-m)})$-competitive algorithm. We will show that using the optimal centers of k-means algorithm gives an approximation to the soft k-means problem. Further, since k-means++ algorithm is $O(\log k)$ competitive to the k-means algorithm, k-means++ is $O(k^{m/(1-m)}) \log k$ competitive algorithm for soft k-means problem.

**Theorem 4.1.** *k-mean centers give an $O(k^{m/(1-m)})$-competitive algorithm for soft k-means.*

The rest of the section provides the proof of this theorem.

Let $c_1^* \cdots c_k^*$ be the optimal k centers of the k-means problem. Then, the objective function of soft k-means with these centers is given by,

$$\begin{aligned}
\Phi(c^*) &\triangleq \sum_{x \in \mathcal{X}} \sum_{i=1}^{k} u_i(x) d(x, c_i^*) \\
&= \sum_{x \in \mathcal{X}} \frac{\sum_{i=1}^{k} d(x, c_i^*)^{-2(1/m-1)}}{\sum_{i=1}^{k} d(x, c_i^*)^{-2/m}} \\
&= \sum_{x \in \mathcal{X}} \frac{\sum_{i=1}^{k} d(x, c_i^*)^{-2(1/m-1)}}{\sum_{i=1}^{k} (d(x, c_i^*)^{-2(1/m-1)})^{\frac{1/m}{1/m-1}}}.
\end{aligned} \tag{4.1}$$

Note that

$$\sum_{i=1}^{k} (d(x, c_i^*)^{-2(1/m-1)})^{\frac{1/m}{1/m-1}} \geq \left(\frac{1}{k}\right)^{\frac{1}{1/m-1}} (\sum_{i=1}^{k} (d(x, c_i^*)^{-2(1/m-1)}))^{\frac{1/m}{1/m-1}}. \tag{4.2}$$

This is because $\sum_{i=1}^{k} a_i^d \geq \frac{1}{k^{d-1}} (\sum_{i=1}^{k} a_i)^d$ for $d \geq 1$. This is true since for a convex function, $f(EX) \leq Ef(X)$. Using uniform discrete distribution over $a_i$ and letting $f(x) = x^d$, we get the above result.

Substituting in (4.1), we get

$$\Phi(c^*) \leq k^{\frac{m}{1-m}} \sum_{x \in \mathcal{X}} (\sum_{i=1}^{k} d(x, c_i^*)^{-2(1/m-1)})^{1 - \frac{1/m}{1/m-1}} \tag{4.3}$$

Substituting $1/m - 1 = g$, we get,

$$
\begin{aligned}
\Phi(c^*) &\leq k^{\frac{m}{1-m}} \sum_{x \in \mathcal{X}} (\sum_{i=1}^{k} d(x, c_i^*)^{-2g})^{-\frac{1}{g}} \\
&\stackrel{(1)}{\leq} k^{\frac{m}{1-m}} \sum_{x \in \mathcal{X}} (\max_{i \in \{1, \cdots, k\}} d(x, c_i^*)^{-2g})^{-\frac{1}{g}} \\
&\stackrel{(2)}{=} k^{\frac{m}{1-m}} \sum_{x \in \mathcal{X}} ((\min_{i \in \{1, \cdots, k\}} d(x, c_i^*))^{-2g})^{-\frac{1}{g}} \\
&= k^{\frac{m}{1-m}} \sum_{x \in \mathcal{X}} \min_{i \in \{1, \cdots, k\}} d(x, c_i^*)^2,
\end{aligned}
\quad (4.4)
$$

where (1) follows since sum of non-negative terms is at-least as large as the maximum of the terms and $g > 0$, and (2) follows since $g > 0$.

Note that this is the objective function of k-means and the centers $c_i^*$ are optimal for this problem. Thus, for any k centers $c_i, 1 \leq i \leq k$, we have

$$
\begin{aligned}
\Phi(c^*) &\leq k^{\frac{m}{1-m}} \sum_{x \in \mathcal{X}} \min_{i \in \{1, \cdots, k\}} d(x, c_i^*)^2 \\
&\leq k^{\frac{m}{1-m}} \sum_{x \in \mathcal{X}} \min_{i \in \{1, \cdots, k\}} d(x, c_i)^2 \\
&= k^{\frac{m}{1-m}} \sum_{x \in \mathcal{X}} \min_{i \in \{1, \cdots, k\}} d(x, c_i)^2 \frac{\sum_{i=1}^{k} d(x, c_i)^{-2/m}}{\sum_{i=1}^{k} d(x, c_i)^{-2/m}} \\
&= k^{\frac{m}{1-m}} \sum_{x \in \mathcal{X}} \frac{\sum_{i=1}^{k} d(x, c_i)^{-2/m} \min_{i \in \{1, \cdots, k\}} d(x, c_i)^2}{\sum_{i=1}^{k} d(x, c_i)^{-2/m}} \\
&\stackrel{(3)}{\leq} k^{\frac{m}{1-m}} \sum_{x \in \mathcal{X}} \frac{\sum_{i=1}^{k} d(x, c_i)^{-2/m} d(x, c_i)^2}{\sum_{i=1}^{k} d(x, c_i)^{-2/m}} \\
&= k^{\frac{m}{1-m}} \sum_{x \in \mathcal{X}} \frac{\sum_{i=1}^{k} d(x, c_i)^{-2(1/m-1)}}{\sum_{i=1}^{k} d(x, c_i)^{-2/m}} \\
&= k^{\frac{m}{1-m}} \Phi(c),
\end{aligned}
\quad (4.5)
$$

where (3) follows since $\min_{i \in \{1, \cdots, k\}} d(x, c_i)^2 \leq d(x, c_i)^2$.

Since the above holds for any centers $c_i$ and thus also for the optimal centers of the soft k-means problem. Thus, we prove that the centers of k-means are atmost $k^{\frac{m}{1-m}}$-competitive to soft k-means.

We can further see that using k-means++ centers give an additional $\log k$ in the approximation. This is because by taking the centers of k-means++ rather than k-means, all the steps till (4.4) directly hold. Also, since k-means++ is $O(\log k)$ competitive, $\sum_{x \in \mathcal{X}} \min_{i \in \{1, \cdots, k\}} d(x, c_i^*)^2 \leq O(\log k) \sum_{x \in \mathcal{X}} \min_{i \in \{1, \cdots, k\}} d(x, c_i)^2$ which adds an extra $O(\log k)$ in the eventual result.

## 5 Streaming Soft k-means

In the previous section, we saw that k-means++ based initialization gives an approximation for soft k-means. This algorithm has been adapted for streaming in the cache register model in [3]. The same algorithm can be used for soft k-means and the result in Theorem 3.1 hold as an approximation to soft k-means. The adapted statement to soft k-means can be stated as follows.

**Lemma 5.1.** *If there is access to memory of size $M = n^\alpha$ for some fixed $\alpha > 0$, then for sufficiently large $n$ the best application of the multi-level scheme described above is obtained by running $r = O(\log(n/M)/\log(M/k \log k))$ levels (which is constant), and choosing the repeated k-means# for all but the last level, in which k-means++ is chosen. The resulting algorithm is a randomized streaming approximation to soft k-means, which is $O(k^{\frac{m}{1-m}} \log k)$-competitive. Its running time is $O(dnk^2 \log n \log k)$.*

This streaming algorithm can also be adapted to streaming over a sliding window when the memory is also limited. In this model, k-means over a window are needed which is moving. In this model, the past data should be removed unlike the cache register model where new data keeps on adding and the old data need not be removed.

Suppose that sliding window length is L and the memory is $O(LK^t(\log k)^t)^{\frac{1}{t+1}}$. Then the cache register model is used with $t+1$ levels with every M points converted to $3k \log k$ centers in the first t steps and M points converted to k points in the last step. Keep the M points at the $t^{th}$ level rather than throwing them away after converting to k points at $t+1$ levels in the cache clustering model.

Shifting window till a length of $L^{\frac{t-1}{1+t}} K^{\frac{2}{1+t}} (3 \log k)^{\frac{2}{1+t}}$ will have impact on only the first $3k \log k$ points in the $t^{th}$ level. So, after this window shift, remove the first $3k \log k$ points at the $t^{th}$ level and add $3k \log k$ points at $t^{th}$ level which are $3k \log k$ centers given by k-mean# algorithm for the M points at $(t-1)^{th}$ level. Thus, we have M points have $t^{th}$ level which give the required k centers. Since at this window shift, the k centers are directly of the sliding window with no extra data point or omitted point, the algorithm is $O(\log k)$ competitive for k-means. If centers are needed at intermediate shift, we can use a weighted average on the last $M - 3k \log(k)$ points completely taken at $t^{th}$ level and weighing first $3k \log k$ points at $t^{th}$ level and the points in the $(t-1)^{th}$ level based on the shift.

**Theorem 5.1.** *The above algorithm with a memory of* $O(L^\epsilon (k \log k)^{1-\epsilon})$ *gives an* $O(\log k)$ *competitive algorithm for k-means for a sliding window of length* L *at every window shift of* $L^{1-2\epsilon}(3k \log k)^{2\epsilon}$.

*Proof.* Using $\epsilon = \frac{1}{t+1}$, we get the above memory requirement. Further, at window shift of $L^{1-2\epsilon}(3k \log k)^{2\epsilon}$, at the $t^{th}$ level, the first $3k \log k$ points will all go out and the new $3k \log k$ points will be added. There is no data point at any but the $(t+1)^{th}$ level and the k centers at the $t+1^{th}$ level are $O(\log k)$ competitive by [3]. □

Since the k-means centers are approximate centers for soft k-means, we can have the following approximation for soft k-means over a sliding window.

**Corollary 1.** *The above algorithm with a memory of* $O(L^\epsilon (k \log k)^{1-\epsilon})$ *gives an* $O(k^{\frac{m}{1-m}} \log k)$ *competitive algorithm for soft k-means for a sliding window of length* L *at every window shift of* $L^{1-2\epsilon}(k \log k)^{2\epsilon}$.

## 6 Empirical Results

In order to evaluate k-means++ initialization in practice, we implemented the two algorithms in Matlab. We label the original algorithm as EM while the one based on k-means++ initialization of centers as EM++. The code is not optimized and is available at [2]. We found that the seeding substantially improves both the running time and the accuracy of EM.

We chose two datasets that are also included in [4]. The first is the *Spam* dataset [24], which consists of 4601 points in 58 dimensions. The second is *Cloud* dataset [10] which consist of 1024 points in 10 dimensions. For each dataset, we tested $k = 10, 25,$ and $50$ and $m = 0.1, 0.25$ and $0.5$.

Since we are testing randomized seeding process, we ran 20 trials on each case. The minimum and average potential and the mean running time are compared between EM and EM++.

We find that seeding gives better speedups as well as better objective functions on these two datasets.

|   |   | Average Φ |   | Mimimum Φ |   | Average T |   |
|---|---|---|---|---|---|---|---|
| m | k | EM | EM++ | EM | EM++ | EM | EM++ |
| 0.1 | 10 | $1.665 \times 10^8$ | 48.08% | $1.016 \times 10^8$ | 23.49% | 34.155 | 37.21% |
| 0.1 | 25 | $1.196 \times 10^8$ | 85.63% | $5.36 \times 10^7$ | 70.88% | 89.256 | 16.87% |
| 0.1 | 50 | $6.304 \times 10^7$ | 89.91% | $7.763 \times 10^6$ | 23.27% | 231.594 | 17.2% |
| 0.25 | 10 | $1.748 \times 10^8$ | 49.63% | $1.076 \times 10^8$ | 22.13% | 37.401 | 28.25% |
| 0.25 | 25 | $8.244 \times 10^7$ | 78.92% | $1.666 \times 10^7$ | 2.01% | 138.632 | 5.52% |
| 0.25 | 50 | $7.838 \times 10^6$ | 16.74% | $6.73 \times 10^6$ | 9.13% | 258.503 | 21.42% |
| 0.5 | 10 | $2.916 \times 10^8$ | 60.81% | $2.329 \times 10^8$ | 51.11% | 72.826 | 57.96% |
| 0.5 | 25 | $4.325 \times 10^7$ | 38.77% | $2.658 \times 10^7$ | 7.37% | 218.268 | 9.63% |
| 0.5 | 50 | $1.202 \times 10^7$ | 7.08% | $1.09 \times 10^7$ | 3.58% | 645.887 | 47% |

Table 1: Experimental results on the *Spam* dataset (n=4601, d=58). For EM, we list the actual potential and time in seconds. For EM++, we list the percentage *improvement* over EM: $100\% \times \left(1 - \frac{\text{EM++value}}{\text{EMvalue}}\right)$.

|     |    | Average Φ |        | Mimimum Φ |        | Average T |          |
| --- | -- | --------- | ------ | --------- | ------ | --------- | -------- |
| m   | k  | EM        | EM++   | EM        | EM++   | EM        | EM++     |
| 0.1 | 10 | 6.462     | 8.74%  | 6.319     | 8.28%  | 3.742     | 24.65%   |
| 0.1 | 25 | 2.403     | 12.27% | 2.226     | 10.09% | 35.456    | 61.02%   |
| 0.1 | 50 | 1.704     | 33.85% | 1.41      | 22.99% | 36.232    | 48.47%   |
| 0.25| 10 | 6.682     | 7.28%  | 6.407     | 6.25%  | 6.401     | 27.04%   |
| 0.25| 25 | 2.318     | 5.37%  | 2.137     | 2.13%  | 28.986    | 36.2%    |
| 0.25| 50 | 1.256     | 4.16%  | 1.197     | 3.7%   | 56.295    | 40.97%   |
| 0.5 | 10 | 8.762     | 10.2%  | 8.762     | 12.75% | 5.142     | 26.45%   |
| 0.5 | 25 | 3.389     | 0.072% | 3.287     | 0%     | 17.087    | -143.93% |
| 0.5 | 50 | 2.339     | 1.65%  | 2.313     | 3.76%  | 90.739    | 18.41%   |

Table 2: Experimental results on the *Cloud* dataset (n=1024, d=10). For EM, we list the actual potential divided by $10^6$ and time in seconds. For EM++, we list the percentage *improvement* over EM: $100\% \times \left(1 - \frac{\text{EM++value}}{\text{EMvalue}}\right)$.